\documentclass[aps, pre, reprint, showpacs,preprintnumbers, a4paper]{revtex4-1}
\usepackage[english]{babel}
\usepackage{graphicx}
\usepackage{dcolumn}
\usepackage{amsmath}
\usepackage{amssymb}
\usepackage{bm}

\begin{document}

    \title{Bifurcation structure of cavity soliton dynamics in a VCSEL with saturable absorber and time-delayed feedback}
    \author{Christian Schelte}
    \email{c.schelte@uib.cat}
    \affiliation{Institute for Theoretical Physics, University of M\"unster, Wilhelm-Klemm-Str.\,\,9, 48149 M\"unster, Germany \\ Departament de F\'isica, Universitat de les Illes Balears, Cra.\,\,de Valldemossa, km 7.5, E-07122 Palma, Spain}
    \author{Krassimir Panajotov}
    \affiliation{Department of Applied Physics and Photonics, B-Phot, Vrije Universiteit Brussel (VUB), Pleinlaan 2, B-1050 Brussels, Belgium}
    \author{Mustapha Tlidi}
    \affiliation{Facult\'{e} des Sciences, Universit\'{e} Libre de Bruxelles, Campus Plaine, C.P. 231, Brussels B-1050, Belgium}
    \author{Svetlana V. Gurevich}
    \email{gurevics@uni-muenster.de}
    \affiliation{Institute for Theoretical Physics, University of M\"unster, Wilhelm-Klemm-Str.\,\,9, 48149 M\"unster, Germany \\ Center for Nonlinear Science (CeNoS), University of M\"unster, Corrensstr.\,2, 48149 M\"unster, Germany\\}
    \date{\today}


\begin{abstract}

    We consider a wide-aperture surface-emitting laser with a saturable absorber section subjected to time-delayed feedback. We adopt the mean-field approach assuming a single longitudinal mode operation of the solitary VCSEL. We investigate cavity soliton dynamics under the effect of time-delayed feedback in a self-imaging configuration where diffraction in the external cavity is negligible. Using bifurcation analysis, direct numerical simulations and numerical path continuation methods, we identify the possible bifurcations and map them in a plane of feedback parameters. We show that for both the homogeneous and localized stationary lasing solutions in one spatial dimension the time-delayed feedback induces complex spatiotemporal dynamics, in particular a period doubling route to chaos, quasiperiodic oscillations and multistability of the stationary solutions.

\end{abstract}

\maketitle


\section{Introduction}

    Cavity Solitons (CSs) are spatially localized light structures in the transverse plane of a nonlinear resonator that result from the balance of nonlinearity and diffraction (for reviews see \cite{Lugiato_po94,Rosanov_po96,Staulinas_book,Kivshar_book,Mandel_04,Ackemann_aamoo09,Tlidi20140101,tlidi2015nonlinear,mihalache2015,he2016dynamics,mihalache2017}). CSs belong to the class of dissipative structures found far from equilibrium, the losses in the system have to be balanced by external energy input. CSs normally require a region in the parameter space where a spatially periodic pattern and a stable homogeneous steady state coexist \cite{Tlidi_prl94,Spinelli_pra98,Bortolozo_njp09}, so that in such a "pinning region" one or more peaks of the patterned state are surrounded by the homogeneous steady state. Recently, Vertical-Cavity Surface-Emitting Lasers (VCSELs) have attracted considerable interest for CS studies and applications because they are inherently made with a short (single longitudinal mode) cavity, which can be transversely quite large \cite{Grabherr_jstqe99}. In the first demonstrations of the existence of CSs in broad-area VCSELs, external coherent light with an appropriate frequency is injected to create the required "pinning region" and CSs have been found both below \cite{Taranenko_pra00,Barland_nature02} and above \cite{Hachair_jstqe06} the lasing threshold. Utilizing the specific polarization properties of VCSELs \cite{Panajotov_aip01}, spatially localized structures have also been created in 40 $\mu m$-diameter VCSELs \cite{Hachair_pra09}, as well as in 80 $\mu m$-diameter VCSELs lasing on a high transverse-order flower mode \cite{Averlant_oe14, Averlant_sr16}. For practical CS applications the need for an external optical injection is a hindrance and a very attractive way to avoid it is the implementation of a saturable absorber in the VCSEL structure \cite{FedorovPRE2000}. Hence, CS properties and dynamics in VCSELs with saturable absorbers have been extensively studied both theoretically \cite{FedorovPRE2000,Bache2005} and experimentally \cite{Genevet_prl08,Elsass_apb10}.

    The impact of time-delayed feedback on CS dynamics has been theoretically investigated for the cases of a driven nonlinear optical resonator \cite{Tlidi_prl09,Gurevich_prl13} and broad-area VCSELs \cite{Panajotov_epjd10,Tlidi_pra12,Pimenov_pra13}. Delayed optical feedback is known to strongly modify the dynamical behavior of semiconductor lasers leading to external cavity mode hopping, periodic or aperiodic dynamics and even coherence collapse \cite{Lenstra_jqe85,Mork_jqe88}. Optical feedback impacts the VCSEL's modal properties and dynamics in quite the same way as those of traditional edge-emitting semiconductor lasers \cite{Chung_ptl91,Besnard_josab99} with the additional peculiarity of introducing polarization switching and two-polarization mode dynamics \cite{Arteaga_jqe06,Arteaga_prl07,Panajotov_jstqe13}. Recently, first studies of CS behavior in optically injected broad-area VCSELs subjected to time-delayed optical feedback have appeared \cite{Panajotov_epjd10,Tlidi_pra12,Pimenov_pra13,Vladimirov_ptrsa14}. These studies elucidated the role of the strength and the phase of the time-delayed feedback for the creation of a drift bifurcation that causes the CSs to spontaneously move. For a saturable absorber VCSEL a period-doubling route to temporal chaos of a single CS has been theoretically predicted for certain feedback parameters \cite{Panajotov_ol14}. More recently, it has been shown that delayed feedback can induce pinning and depinning of cavity solitons when the resonator is illuminated by an inhomogeneous spatial gaussian pumping beam \cite{Tabbert_pre17}. It has to be noted that time-delayed feedback in spatially extended complex systems has a broader relevance than just laser physics and nonlinear optics. It concerns all fields of natural science~\cite{scholl2008handbook}, for instance, chemical reaction-diffusion systems \cite{Kyrychko2009, tlidi2013delayed,GurevichRD,gurevich2014time}.

    Previously, oscillatory dynamics of CSs have been observed in systems without optical feedback: in the Lugiato-Lefever model \cite{Lugiato_prl87} of a driven optical nonlinear cavity \cite{Stroggie_csf94,Firth_josab02,Turaev_prl12} and in a model of a VCSEL with a saturable absorber extended beyond the mean-field approximation \cite{Columbo_epjd10}. Furthermore, a period doubling route to chaos has been predicted for localized structures in the Lugiato-Lefever equation \cite{Nozaki_pd86,Barashenkov_pre11} in a forced and damped van der Pol model \cite{Martinez_epl99}. Spatiotemporal chaos has also been reported for the Lugiato-Lefever equation \cite{liu}. Experimentally, such oscillatory dynamics of localized structures have been observed in an optically pumped VCSEL with a saturable absorber \cite{Elsass_epjd10}. More recently, two-dimensional dissipative optical rogue waves have been predicted to occur in VCSEls with delayed feedback \cite{Tlidi_Krass_17} or without \cite{PhysRevA.95.023841}.

    In this paper we carry out a detailed investigation of the bifurcation structure of time-delayed feedback induced complex dynamics in a VCSEL with a saturable absorber. Using bifurcation analysis, direct numerical simulations and numerical path continuation methods we show that the feedback impacts the homogeneous lasing solution and the localized CS solutions in a similar way, causing oscillatory dynamics with either a period doubling or quasiperiodic route to chaos, as well as multistability of the stationary solutions. The paper is organized as follows: In section II we introduce the mean-field model of broad-area VCSEL with a saturable absorber and time-delayed feedback. In section III we discuss its stationary spatially homogeneous and localized solutions with an emphasis on the impact of the feedback parameters on the branches of stationary solutions. In section IV we reveal the underlying bifurcation structure and in section VI we carry out direct numerical simulations to see the dynamical behavior under time-delayed feedback. In section VI we perform numerical path continuation calculations and map the bifurcation structure of the system in a plane of feedback parameters. Finally, we conclude in section VII.


\section{Model system}

    We investigate a model system for a wide aperture semiconductor laser, specifically a Vertical Cavity Surface Emitting Laser (VCSEL) that consists of a gain section and a saturable absorber section sandwiched between two Distributed Bragg Reflectors (DBRs) that form the optical cavity. It is subjected to time-delayed optical feedback by forming an external cavity using a distant mirror to reflect the field. We adopt the Rosanov \cite{rozanov1975kinetics} and Lang-Kobayashi \cite{lang1980external} approximation to model the delayed feedback. In this approximation, the feedback field is sufficiently attenuated, and it can be modeled by a single delay term with a spatially homogeneous coefficient. In addition, we assume that the laser operates in a single-longitudinal mode. The system of model equations reads \cite{Panajotov_epjd10,Tlidi_pra12,Pimenov_pra13,Panajotov_ol14}
    \begin{eqnarray}
        \partial_t E &=& \left[(1-i\,\alpha)\,N+(1-i\,\beta)\,n-1+i\,\nabla_\perp^2\right]\,E+ \nonumber \\
            &+&\eta\,e^{i\,\varphi}\,E(t-\tau) \label{eq:model1}\\
        \label{eq:model2}
        \partial_t N &=& b_1\left[\mu-N\,(1+|E|^2)\right] \\
        \label{eq:model3}
        \partial_t n &=& b_2\left[-\gamma-n\,(1+s\,|E|^2)\right]\,
    \end{eqnarray}
    where $E=E(r,t)$\,, $r=(x,y)$ is the slowly varying electromagnetic field envelope and $N=N(r,t)$ ($n=n(r,t)$) measures the state inversion of the carriers in the gain (absorber) section. Time is scaled to the photon lifetime and space is scaled to the diffraction length. Here, $b_1$ ($b_2$) is the ratio of the gain (absorber) carrier lifetime to the photon lifetime, $\mu$ is the gain current and $\gamma$ the absorber voltage. Furthermore, $\alpha$ ($\beta$) is the linewidth enhancement factor of the gain (absorber) section and $s$ is the ratio of the saturation intensities of the gain and absorber. Finally, $\eta$ is the relative strength of the time delayed feedback, $\tau = 2L_{ext}/c$ is the delay time with $c$ the speed of light and $L_{ext}$ the external cavity length and $\varphi$ is a delay phase parameter that describes a phase shift on the time scale of the fundamental lasing frequency, due to, e.g., moving the mirror for a distance shorter than the wavelength.

    Equations~\eqref{eq:model1}-\eqref{eq:model3} have two types of homogeneous steady-state solutions. The trivial off solution reads \cite{Bache2005}
    \begin{equation}
        E = 0\,, \qquad N = \mu\,, \qquad n = -\gamma\,.
    \end{equation}
    It becomes unstable at the lasing threshold $\mu_\mathrm{th}=1+\gamma$\,. From this point, a non-trivial branch of spatially homogeneous lasing solutions (so-called continuous waves (CWs)) emerges
    \begin{equation}
        E = |E|e^{i\omega t}, \quad N = \dfrac{\mu}{1+|E|^2}, \quad n = \dfrac{-\gamma}{1+s|E|^2},
    \end{equation}
    where $\omega$ is the frequency shift of the slowly varying field envelope.

    Notice that for a stable lasing solution the intensity must be large enough to overcome the saturable absorber which causes the CW branch to initially be unstable and lean towards lower gain. It then folds in a saddle-node bifurcation at
    \begin{equation}
        \mu_\mathrm{fold}=\dfrac{\left(\sqrt{s-1}+\sqrt{\gamma}\right)^2}{s}\,.
    \end{equation}
    This gives rise to a regime below the lasing threshold where stable lasing solutions coexist with a stable off solution. Note that the time delayed feedback can shift both bifurcation points.

    The chosen parameter values for this article reflect an experimental setup. Unless specified otherwise they are
    \begin{gather*}
        \alpha=2\,,\,\, \beta=0\,,\,\, b_1=0.04\,,\,\, b_2=0.02\,,\,\, s=10\,, \\
        \mu=1.42\,,\,\, \gamma=0.5\,,\,\, \tau=100\,.
    \end{gather*}


\section{Stationary solutions}

    First we analyze the stationary solutions of the system \eqref{eq:model1}-\eqref{eq:model3} by purely analytical means and standard path continuation techniques. Stationary here means the field profile is constant in time while it rotates uniformly in the complex phase.


\subsection{Continuous waves}

\begin{figure} [h]
    \includegraphics[scale=.5]{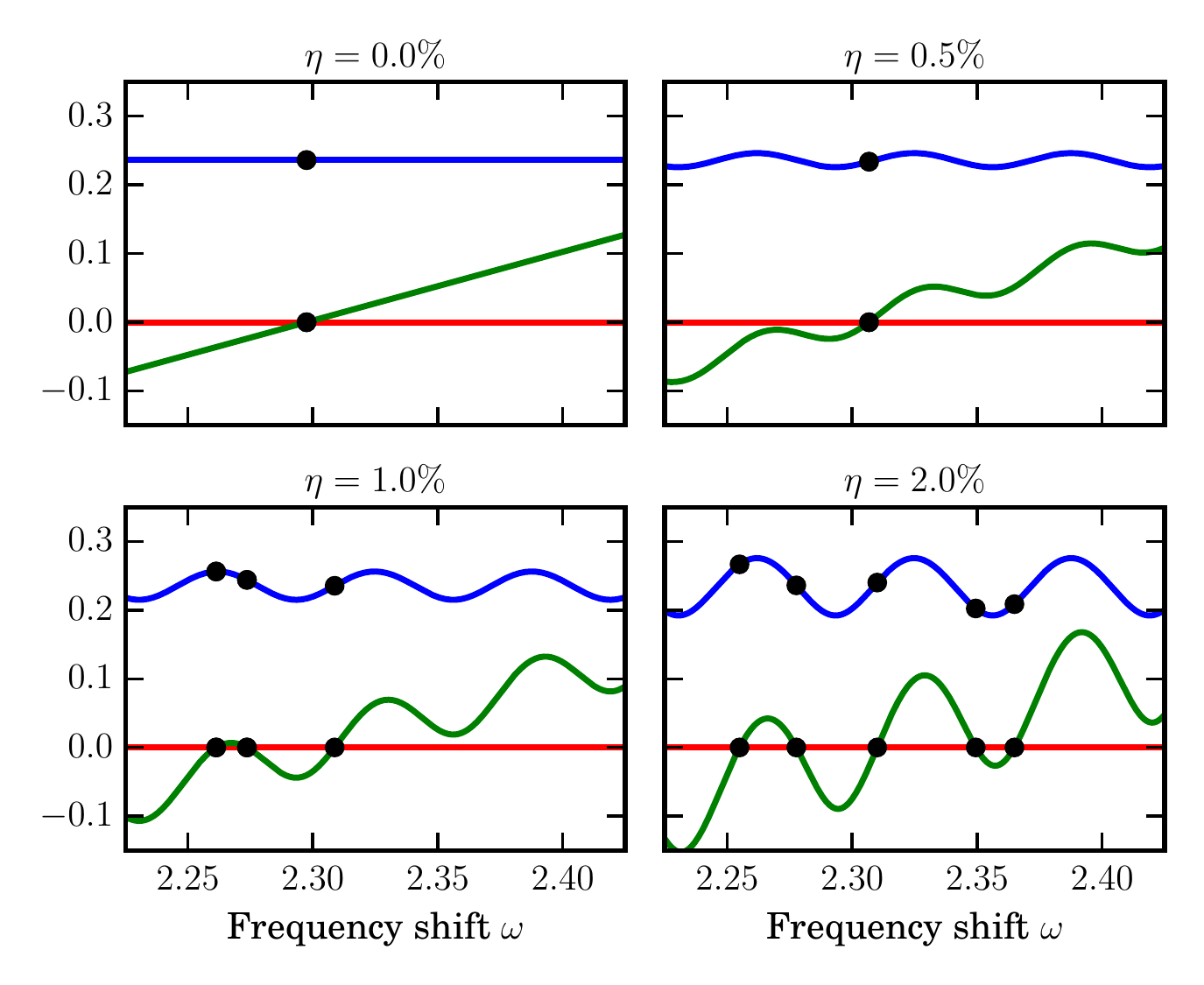}
        \caption{(color online)
            A graphical scheme to determine CW solutions of Eqs.~\eqref{eq:model1}-\eqref{eq:model3}. The intensity $I$ as a function of the frequency shift $\omega$ is shown in blue from solving Eq.~\eqref{eq:graphical_b}. The r.h.s. of Eq.~\eqref{eq:graphical_a} as a function of $\omega$ is shown in green and the l.h.s. in red. Where these lines intersect a solution exists, indicated with black dots. Increasing delay causes stronger oscillations in the r.h.s. curve, inducing a series of saddle-node bifurcations as the l.h.s. is crossed at additional points. The corresponding intensities can be read from the intensity curve. Here $\varphi=0$\,.
        }
        \label{fig:graphical}
    \end{figure}

    The stationary CW solutions take the form
    \begin{equation}
        E = |E|\,e^{i(kx-\omega t)}\,, \quad \partial_t |E| = 0\,,
    \end{equation}
    with the wavenumber $k$ and the frequency shift $\omega$ and the corresponding carrier densities
    \begin{equation}
        N = \frac{\mu}{1+I}\,, \quad n = \frac{-\gamma}{1+sI\,}\,,
    \end{equation}
    with the field intensity $I=|E|^2$\,. For a stationary field amplitude the delayed field is merely shifted in phase $E(t-\tau) = e^{i\omega\tau}E(t)$\,. The model equations \eqref{eq:model1}-\eqref{eq:model3} then simplify to
    \begin{eqnarray}
        0 = \dfrac{\mu\,(1-i\,\alpha)}{1+I}\,-\,\dfrac{\gamma\,(1-i\,\beta)}{1+s\,I}-1+\\ \nonumber
        +i(\omega-k^2)+\,\eta\,e^{i(\omega\tau+\varphi)}.
    \end{eqnarray}
    Separating the real and imaginary parts we have
    \begin{subequations}
    \begin{align}
        \label{eq:graphical_a}
        k^2 &= \omega-\dfrac{\alpha\mu}{1+I}+\dfrac{\beta\gamma}{1+sI}+\eta\,\mathrm{sin}(\omega\tau+\varphi)\,, \\
        \label{eq:graphical_b}
        0 &= \dfrac{\mu}{1+I}-\dfrac{\gamma}{1+sI}-1+\eta\,\mathrm{cos}(\omega\tau+\varphi)\,.
    \end{align}
    \end{subequations}
    With this the solutions can be found graphically, see Fig.~\ref{fig:graphical}. The second equation yields all possible intensities $I$ as a function of $\omega$ (cf. Fig.~\ref{fig:graphical} upper line in blue). With $I$ we can calculate the right hand side of the first equation (inclined line in green). Exact solutions (black dots) are found wherever this line intersects with the lower flat line in red at the value of $k^2$\,. For increasing values of $\eta$ additional solutions appear in a series of saddle-node bifurcations.

    Note that linear stability analysis of the CWs shows modulation instability for all wave numbers $k$ for the investigated domain size.

\subsection{Cavity solitons}

    To find a branch of one-dimensional cavity solitons (CSs) we assume a stationary complex profile $A(x)$ of the field envelope that rotates with a constant frequency shift $\omega$ like the CW solutions~\cite{FedorovPRE2000}:
    \begin{equation}
        E(x,t) = A(x)\,e^{-i\omega t}\,.
    \end{equation}
    This profile consists of an amplitude profile $a(x)$ and a phase profile $\varphi(x)$\,.
    \begin{subequations}
    \begin{gather}
        A(x) = a(x) e^{i\varphi(x)}\,, \\
        \quad q = \partial_x \varphi\,, \\
        \quad k = \dfrac{1}{a}\partial_x a\,, \\
        f(|A|^2) = \dfrac{(1-i\alpha)\mu}{1+|A^2|} - \dfrac{(1-i\beta)\gamma}{1+s|A^2|} -1\,.
    \end{gather}
    \end{subequations}
    Because the whole system is phase invariant, only the derivative of the phase profile is important, reducing the number of necessary variables to three. With this we can write the system in the form
    \begin{subequations}
    \begin{align}
        \partial_x a &= a k\,, \\
        \partial_x q &= -2 qk + \operatorname{Re}[f(a^2)] + \eta\,\mathrm{cos}(\omega\tau+\varphi)\,, \\
        \partial_x k &= -\omega + q^2 - k^2 - \operatorname{Im}[f(a^2)] - \eta\,\mathrm{sin}(\omega\tau+\varphi)\,,
    \end{align}
    \end{subequations}
    following \cite{1063-7818-27-11-A04}. We can now treat it as a boundary value problem and apply standard path-continuation packages like, e.g., AUTO-07P \cite{DoKK1991ijbc,DoKK1991ijbcb} to obtain a branch of stationary solutions.


\subsection{Effective phase}

    To get an alternative view at the influence of the time-delayed feedback on the solution structure we introduce an effective phase parameter $\vartheta$~\cite{PuzyrevPRA16}:
    \begin{equation} \label{eq:effective_phase}
        \quad \vartheta = (\omega\,\tau + \varphi) \, \mathrm{mod} \, 2\pi\,.
    \end{equation}
    This yields the branches of localized solutions that have the same angle of interference with the delayed field. They form a tube shaped manifold of all possible solutions of the system for a given delay strength. This view is not directly accessible experimentally. For any combination of $\tau$ and $\varphi$ one can also get the actual branches by solving equation \eqref{eq:effective_phase} implicitly. In particular, it can be shown that the number of multistable solutions grows linearly with $\tau$~\cite{PuzyrevPRA16}.


\subsection{Solution structure}

\begin{figure} [h]
        \includegraphics[scale=.5]{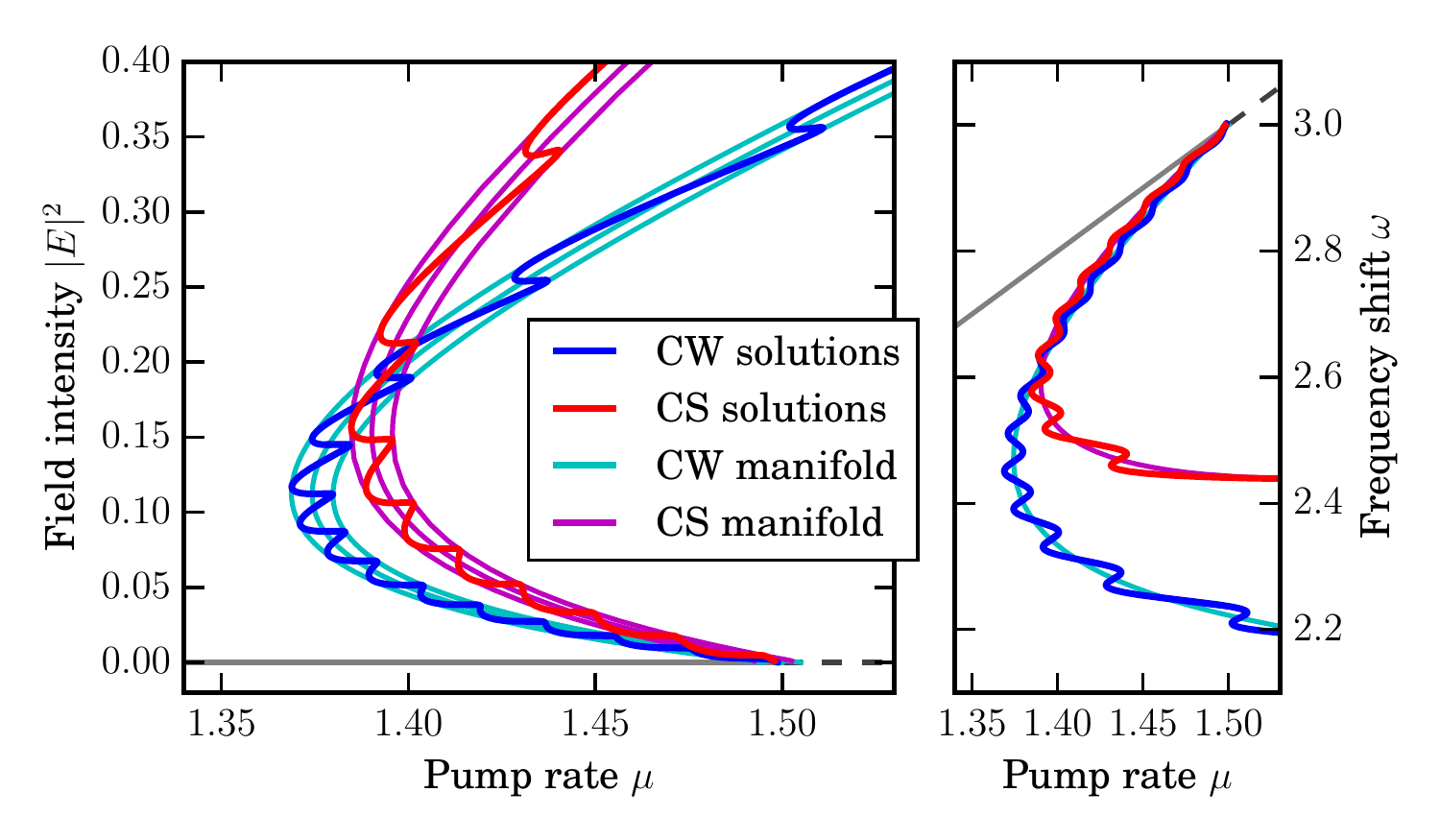}
        \caption{(color online)
            Solution structure of the stationary solutions of Eqs.~\eqref{eq:model1}-\eqref{eq:model3}. Left (right) panel shows the intensities $|E|^2$ (frequency shifts $\omega$) as a function of the gain $\mu$\,. Actual solutions for a fixed delay time $\tau=100$ and delay phase $\varphi=0$ are drawn in blue (red) for CWs (CSs). Both exhibit a similar snaking shape in both $|E|^2$ and $\omega$ due to a series of saddle-node bifurcations induced by time-delayed feedback. For reference, the central C-shaped curves in cyan (magenta) for CWs (CSs) show the solutions without delay. Changing $\varphi$ moves the snaking curve periodically along the tube-shaped manifold of solutions.
        }
        \label{fig:CS_eta05}
    \end{figure}

    Figure \ref{fig:CS_eta05} shows the obtained solution structures of CWs (cyan and blue) and CSs (magenta and red) for $\eta=0.5\%$ and $\varphi=0$ with $\mu$ as the control parameter. The left (right) panel shows the intensities (frequency shifts) of the corresponding solutions. For CSs the intensity in the center is shown. Note that both the CWs and CSs have a very similar solution structure.

    The C-shaped curves represent the solution manifolds. Any point between the outer C-curves is a solution for appropriate delay parameters. The left (right) curves represent fully constructive (destructive) interference, i.e., $\vartheta=0$ ( $\vartheta=\pi$)\,. The central curves represent the solution branch without delay. In the presence of delay this curve effectively gets shifted in $\mu$ as the feedback either helps or hinders the field in the cavity. Aside from the shift the changes to the curves are minimal.

    The snaking curves show the actual solution branches for $\tau=100$ and $\varphi=0$\,. They form through a series of saddle node bifurcations induced by the delayed feedback. Along the branches the stability of the stationary solutions alternates. For the intensities the positive (negative) slopes are stable (unstable) and vice versa for the corresponding frequency shifts. Note that a similar multistability effect was experimentally observed in a broad-area VCSEL with frequency-selective feedback~\cite{TanguyPRA2006}.

    An animation of Fig.~\ref{fig:CS_eta05} showing the effect of the feedback parameters on the solution structure is available in the Supplemental Material \footnote{See Supplemental Material at [URL will be inserted by publisher] for an animation of the stationary solution structure seen in Fig.~\ref{fig:CS_eta05} showing the effect of the feedback parameters.}.

\section{Phase bifurcation and mustistability}

\begin{figure} [h]
        \includegraphics[scale=.5]{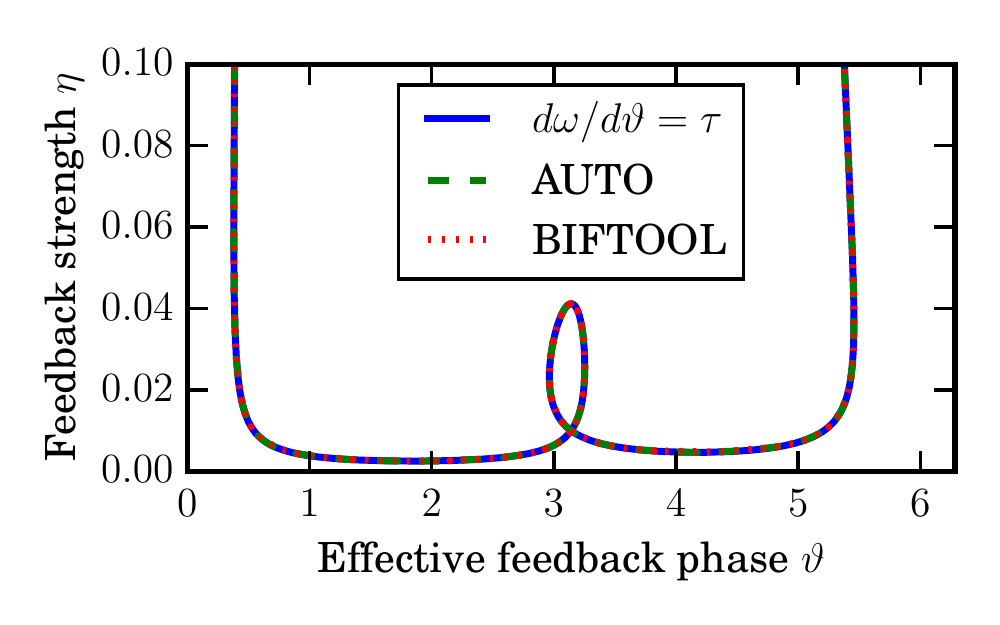}
        \caption{(color online)
            A bifurcation diagram for the folds with the delay strength $\eta$ and effective phase $\vartheta$ as control parameters. The solid blue line shows a branch of solutions that fulfill the condition~\eqref{eq:dwdt} for a phase bifurcation. The dashed green (dotted red) shows the fold continuation in AUTO-07P (DDE-BIFTOOL). The two continuations yield equivalent results. The folds can be attributed to a phase instability induced by the time-delayed feedback.
        }
        \label{fig:dwdt}
    \end{figure}

    Due to translational and phase-shift symmetries of Eqs.~\eqref{eq:model1}-\eqref{eq:model3}, the point spectrum of the corresponding one-dimensional linear eigenvalue problem has two zero eigenvalues corresponding to the even phase-shift neutral mode and the odd translational neutral mode. Drift or phase bifurcations can occur when the eigenvalue of the corresponding neutral mode $\boldsymbol{\psi_0}$ becomes doubly degenerate with geometrical multiplicity one. There, the critical real eigenvalue passes through zero at the bifurcation point, so that the corresponding critical eigenfunction at this point is proportional to the neutral mode. This critical eigenvalue can either be a delay-induced branch of zero eigenvalue or correspond to a Galilean mode, which is generally nonzero due to delayed feedback~\cite{1063-7818-27-11-A04, PuzyrevPRA16}. In \cite{PuzyrevPRA16} a general expression for the onset of drift and phase bifurcations was derived:
    \begin{equation}
        \eta\tau=-\frac{<\boldsymbol{\psi}^{\dag}_0|\boldsymbol{\psi}_0>}{<\boldsymbol{\psi}^{\dag}_0|\mathrm{B}|\boldsymbol{\psi}_0>}\,,
    \end{equation}
    where $\boldsymbol{\psi_0}$ is a neutral eigenfunction, $\boldsymbol{\psi^{\dag}_0}$ the corresponding adjoint eigenfunction and $\mathrm{B}$ is the rotation matrix describing the phase shift due to the delay. Note that both drift- and phase-bifurcation thresholds tend to zero in the limit of large delays. While in the case of the drift bifurcation a pitchfork bifurcation takes place, the phase bifurcation corresponds to a saddle-node bifurcation where a pair of solutions merge and disappear. Note that this fold condition follows directly from Eq.~\eqref{eq:effective_phase} and can be written as 
    \begin{equation} \label{eq:dwdt}
        \dfrac{d\omega}{d\vartheta}=\dfrac{1}{\tau}\,.
    \end{equation}
    Figure \ref{fig:dwdt} shows the branch of stationary localized solutions satisfying the fold condition~\eqref{eq:dwdt} (solid blue line) along with results from a fold continuations performed in AUTO-07P (dashed green line) and in DDE-BIFTOOL (dotted red line). One can see that all three calculations yield the same result. This demonstrates the equivalence of the continuations and identifies a phase bifurcation as the cause of the delay-induced multistability.

\section{Direct numerical simulations}

    \begin{figure} [b]
        \includegraphics[scale=.5]{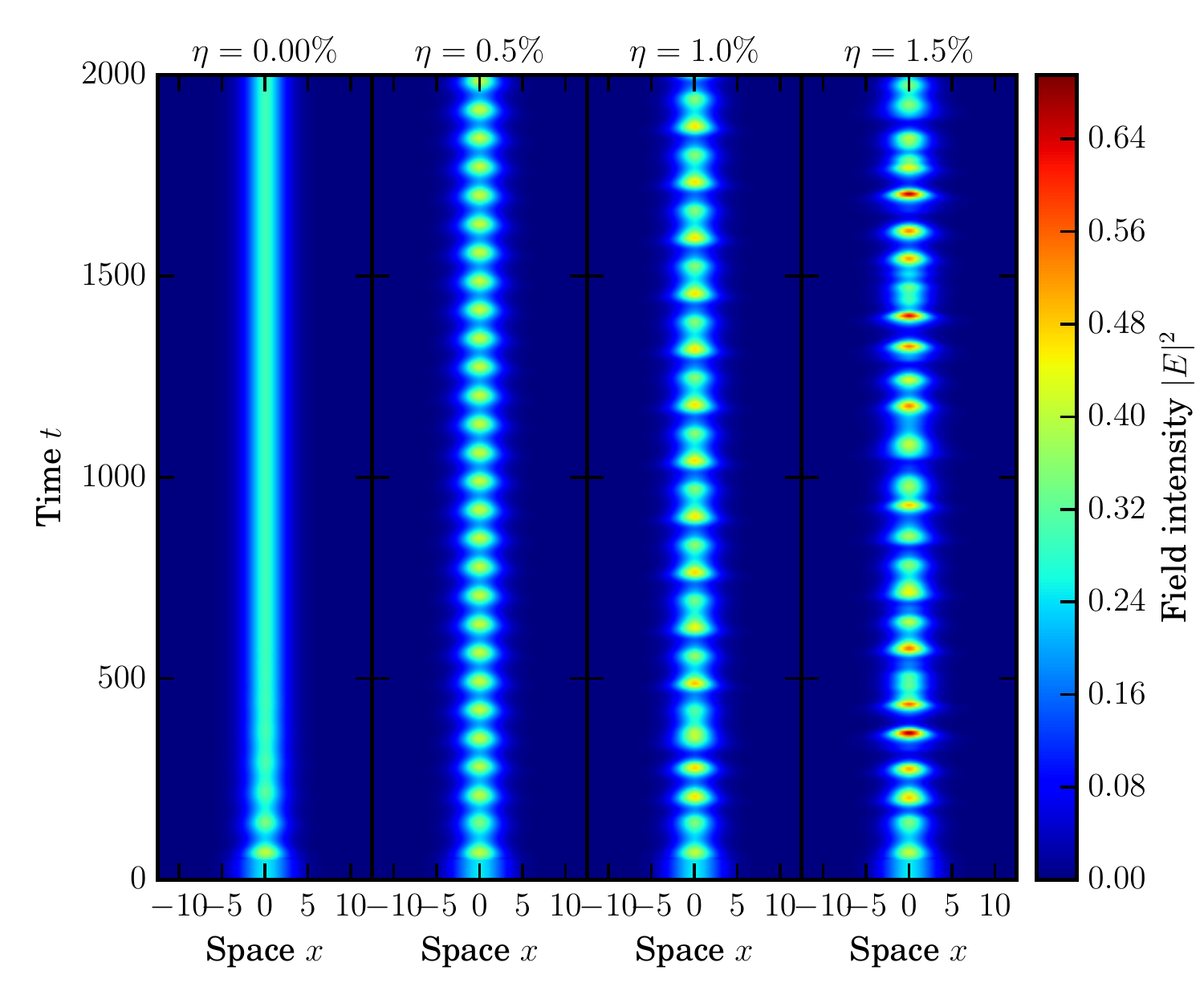}
        \caption{(color online)
            Space-time plots of one-dimensional simulations of Eqs.~\eqref{eq:model1}-\eqref{eq:model3} calculated for a delay time $\tau=100$\,, delay phase $\varphi=0$ and various delay strengths $\eta$\,. Without delay the system equilibrates quickly to a stationary lasing localized structure. With increasing $\eta$ it becomes Hopf unstable and oscillates in intensity. Further increasing $\eta$ causes the periodic orbit to undergo a series of period doubling bifurcations that leads into chaos. The rightmost panel shows the chaotic behavior that is characterized by strong intensity spikes.
        }
        \label{fig:spacetime}
    \end{figure}

    In addition to the drift bifurcation leading to traveling CSs~\cite{FedorovPRE2000,Prati2010,PuzyrevPRA16} and the phase bifurcation giving rise to the multistability of CSs solutions, time-delayed feedback can also induce Andronov-Hopf bifurcations. Indeed, in~\cite{Panajotov_ol14} it has been shown that the inclusion of the feedback term leads to the formation of breathing CSs with a period doubling route to chaos. In order to analyze transitions between different oscillating solutions of a single CS, one-dimensional direct numerical simulations of the system have been performed using the classical Runge-Kutta method on an equidistant mesh combined with a pseudospectral method for spatial derivatives. Note that interpolation of the delay term to reach the same order of convergence as the time stepping scheme is not needed in this case because the delay strength is small. Figure \ref{fig:spacetime} shows exemplary space-time plots of the field intensity for different values of $\eta$ at $\varphi=0.5\,\pi$\,. Without delayed feedback the system forms localized lasing structures with a steady intensity profile. Increasing the feedback causes an Andronov-Hopf bifurcation and the intensity continuously oscillates. These oscillations undergo a period doubling bifurcation leading to chaos as demonstrated in \cite{Panajotov_ol14}. In the right panel one can see irregular oscillations with strong spikes of intensity that are otherwise never achieved in this system. This identifies the behavior as chaotic in distinction to quasi-periodic oscillations that would have a clearly limited interval of intensity. Indeed, for certain delay parameters there is also a torus bifurcation which, however, is not explicitly represented in this figure. 

\begin{figure} [b]
        \begin{tabular}{ll}
            (a)\\
            \includegraphics[scale=.5]{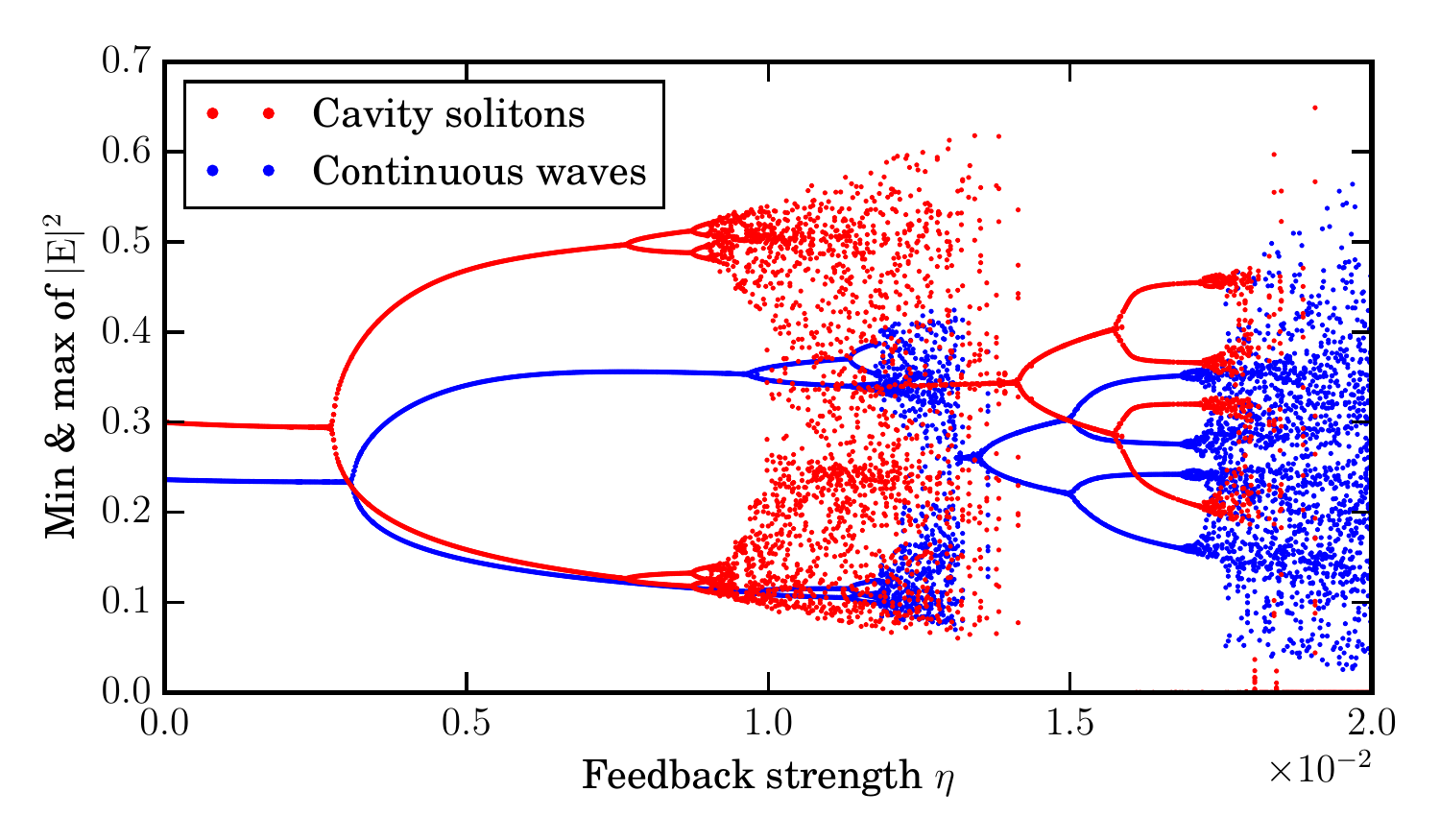}\\
            (b)\\
            \includegraphics[scale=.5]{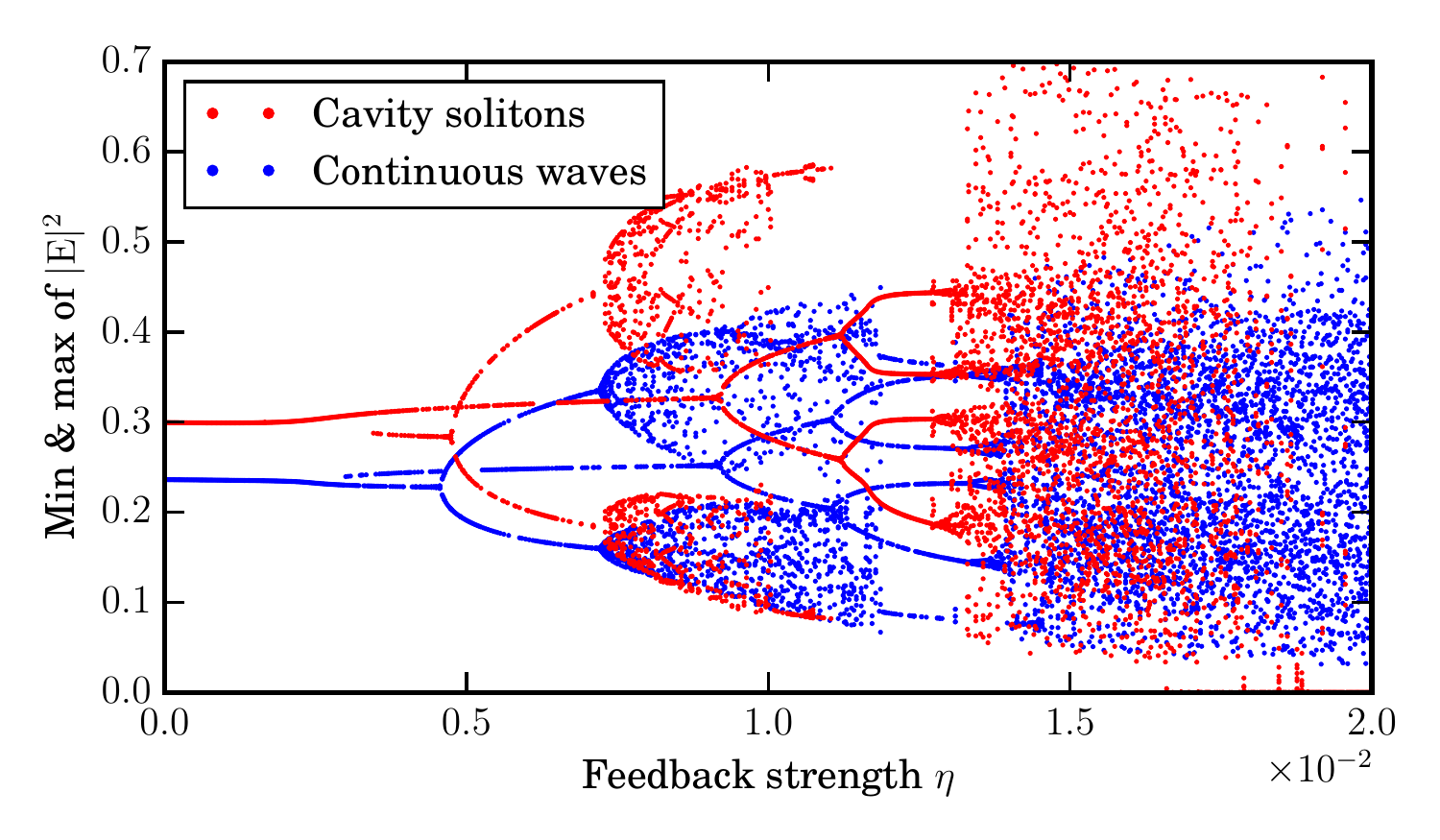}
        \end{tabular}
        \caption{(color online)
            Bifurcation diagram with the delay strength $\eta$ as the control parameter calculated for (a) $\varphi=0$\,, (b) $\varphi=1.48\pi$\,. In direct numerical simulations a time series is analyzed after the system has had sufficient time to settle. The intensity extrema of the time series are shown as blue (red) dots for the CW (CS) case. For increasing $\eta$ we see either a period doubling or quasiperiodic route to chaos. The bifurcation points move with changing delay phase $\varphi$\,. In a range of $\varphi$ there exist two separate windows with a route to chaos. These are actually connected through $\varphi$\,, i.e., the window appears below the first saddle-node bifurcation on the lower branch and later moves off to the right for increasing $\varphi$\,. Due to the periodicity of $\varphi$ a new window appears before the other one vanishes.
        }
        \label{fig:FB}
    \end{figure}

    To characterize the different kinds of temporal behavior we trace the extrema of the intensity field in time. For CSs the intensity value in the center was traced. Plotting these extrema after the system has reached a final state of operation yields a bifurcation diagram with $\eta$ as the control parameter. Figure \ref{fig:FB}~(a) shows an exemplary bifurcation diagram for $\varphi=0$\,. Note that several windows of stationary behavior and chaotic dynamics can be observed. The location of these windows moves with the feedback phase $\varphi$ and can be associated with the aforementioned delay-induced multistability of the stationary CS solution. In particular, at $\varphi\approx\pi$ the period doubling route of the lower branch starts to appear in coexistence with the upper branch. For increasing $\varphi$ the bifurcation points move towards lower $\eta$\,. At the saddle node bifurcation at $\varphi\approx1.5\,\pi$ they change direction and continue towards higher $\eta$\, \textemdash~now as the upper branch. Here, the overlap of the two period doubling routes is most pronounced (see Fig. \ref{fig:FB}~(b), where an exemplary bifurcation diagram for $\varphi=1.48\,\pi$ is presented). An animation showing the dynamics of the bifurcation diagram as one changes the feedback phase is available in the Supplemental Material \footnote{See Supplemental Material at [URL will be inserted by publisher] for an animation of the bifurcation diagrams in Fig.~\ref{fig:FB} showing the effect of the feedback phase $\varphi$\,.}.

    The obtained bifurcation diagram for the homogeneous lasing solution resembles the diagram for CSs in appearance. In particular, Andronov-Hopf bifurcations as well as saddle-node bifurcations of CSs and of the homogeneous lasing solution occur at similar but not exactly the same values of $\eta$\,. That is, the dynamical behavior of the homogeneous lasing solutions can be first studied in detail using, e.g., standard path continuation tools for delay differential equations \cite{Engelborghs:2002:NBA:513001.513002} as this analysis is much simpler than the complete analysis of the spatially distributed problem.

\section{Delay continuation}

    \begin{figure} [b]
        \includegraphics[scale=.5]{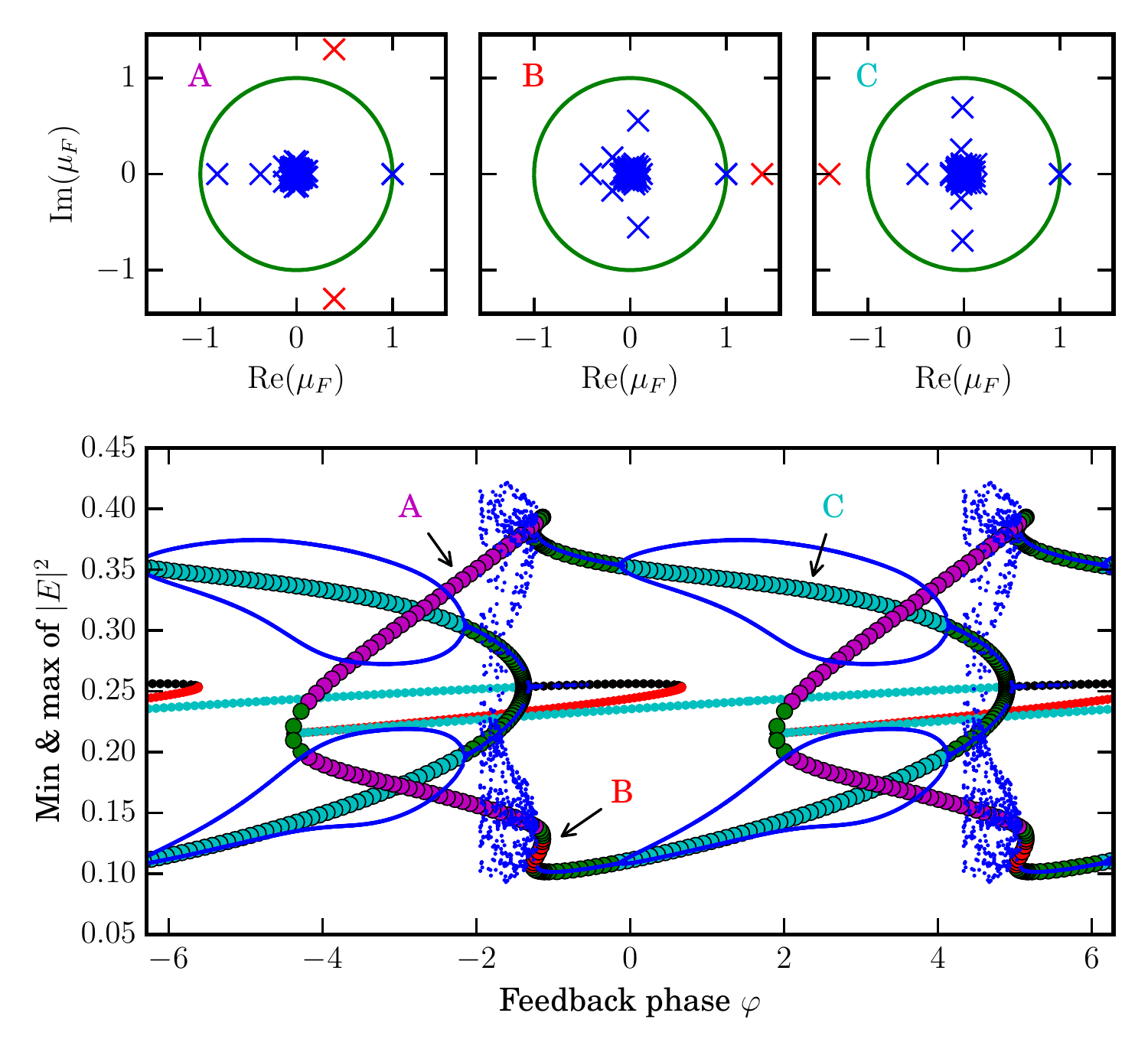}
        \caption{(color online)
            The lower panel shows a bifurcation diagram obtained in DDE-biftool with the delay phase $\varphi$ as the control parameter. Stationary solutions are drawn as dots colored black for stable, red for the unstable connection between the folds and cyan for Andronov-Hopf instability. The periodic orbits connecting the Andronov-Hopf bifurcations are shown as filled circles in green for stable, cyan for period doubling, red for cyclic-fold and magenta for torus bifurcation. Corresponding data from direct numerical simulations is plotted as small blue dots for comparison. The simulations are in agreement with the continuation. \\ The upper panels show the Floquet multipliers corresponding to the unstable periodic orbits. Values outside the unit circle indicate the respective instabilities, i.e., real and larger than one means cyclic-fold, real and smaller than minus one means period doubling and a complex pair with an absolute value larger than one means torus bifurcation.
        }
        \label{fig:biftool}
    \end{figure}

    We use DDE-BIFTOOL with the extensions for periodic orbits and rotational symmetry to analyze the dynamic solutions of the system. DDE-BIFTOOL \cite{Engelborghs:2002:NBA:513001.513002} is a path continuation toolbox for delay differential equations (DDEs) in Matlab. Since DDE-BIFTOOL is designed to continuate delay differential equations, Eqs.~\eqref{eq:model1}-\eqref{eq:model3} can be approximated by a set of coupled delay differential equations. However, the underlying algorithms' execution times scale badly with the system dimension. In particular, calculations for a single equation in space have shown an effective limit to spatial resolution of 64 mesh points on contemporary desktop hardware~\cite{Tabbert_pre17}. For the 4d system at interest we estimate a limit of only 16 meshpoints in space which is hardly sufficient. We therefore look only at the CW case since CSs are expected to behave similarly as was demonstrated before.

    Figure \ref{fig:biftool} shows the results of the analysis with DDE-BIFTTOL for $\eta=1\%$ on the full interval of $\varphi$\, in the lower panel. The stationary solution is stable (black dots) only for a small interval of $\varphi$\, while most of it is Hopf unstable (cyan dots). Between the two folds the branch is unstable (red dots). The periodic orbits are represented by circles (green when stable) at the extrema of their intensity profiles in time. They connect the Andronov-Hopf bifurcations. On two separate intervals they are unstable to period doubling (green circles) and a torus bifurcation (magenta circles), respectively. There is also a cyclic fold before the torus bifurcation with the unstable part in red. For comparison the results from direct numerical simulations are shown as blue dots. Both the continuation and the simulations are in good agreement.

    The upper panels of Fig.~\ref{fig:biftool} show the respective Floquet multipliers of representative unstable periodic orbits. In the left panel the torus bifurcation is identified by a complex pair of Floquet multipliers outside the unit circle. In the middle panel the cyclic-fold is identified by a real Floquet multiplier larger than one. In the right panel the period doubling is identified by a real Floquet multiplier smaller than minus one.

\begin{figure} 
        \includegraphics[scale=.5]{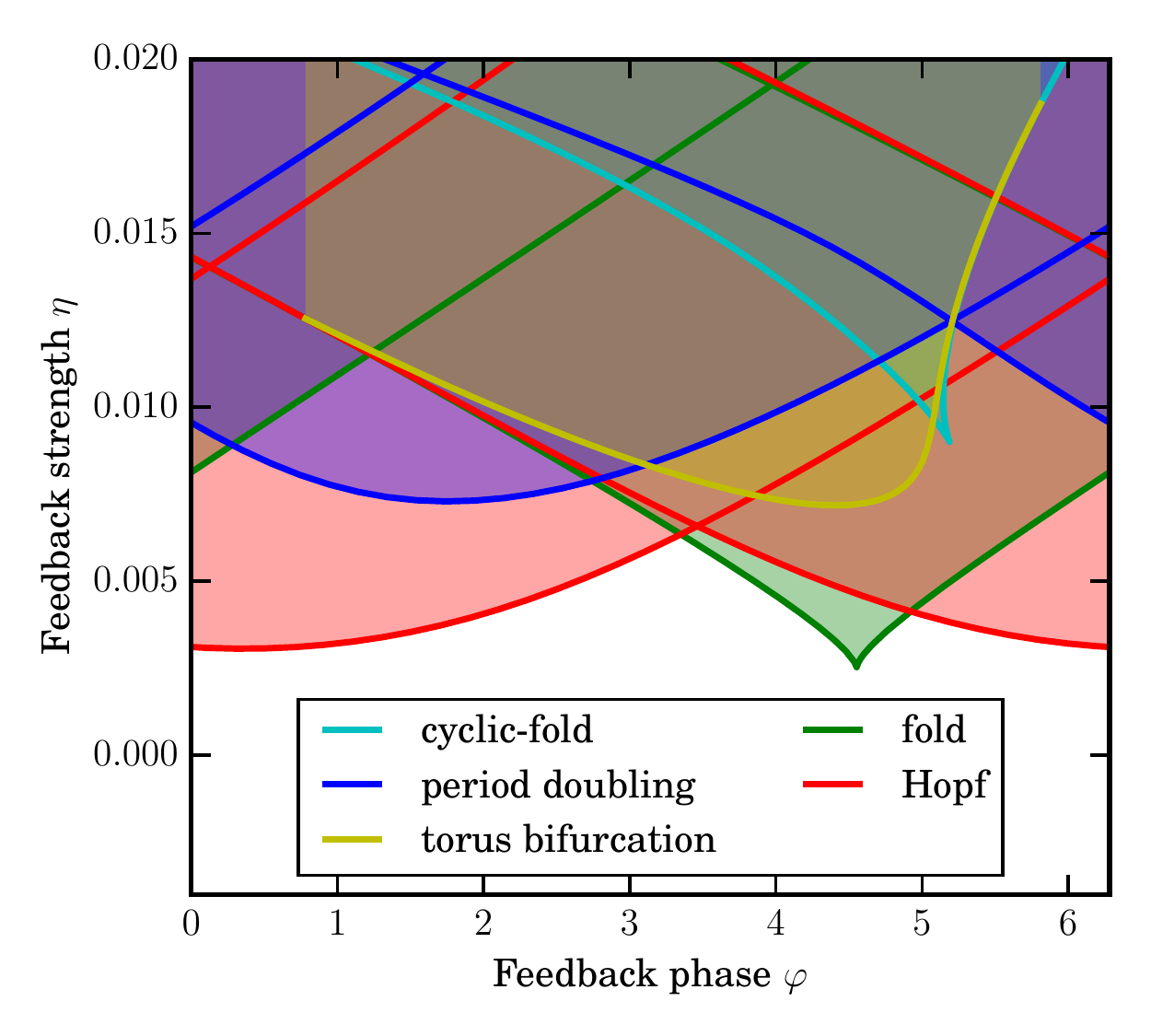}
        \caption{(color online)
            Bifurcation diagram for CWs obtained with DDE-biftool with the feedback strength $\eta$ and feedback phase $\varphi$ as control parameters. The folds (Hopf thresholds) of the stationary solutions are drawn in green (red). The cyclic folds of the periodic orbits are shown in cyan and the period doubling (torus) thresholds in blue (yellow). Areas are colored corresponding to the various instabilities present.
        }
        \label{fig:patches}
    \end{figure}

    Finally, Fig.~\ref{fig:patches} shows the full bifurcation diagram for CW stationary solutions and their periodic orbits in the ($\varphi, \eta$)-plane. For the stationary solutions the Andronov-Hopf bifurcation is shown in red and the saddle-node bifurcation in green. For the periodic orbits the period doubling bifurcation is shown in blue, the cyclic-folds in cyan and the torus bifurcation in yellow. The torus bifurcation branch connects the cyclic-fold with the crossing point of the stationary fold with the Andronov-Hopf bifurcation. The colored areas represent the respective combination of delay-induced instabilities. One can see that increasing $\eta$ leads to complex spatio-temporal behavior of the homogeneous lasing solution including multistability and coexistence of stationary states with periodic and aperiodic dynamics.

\section{Conclusion}

    To conclude, we carried out a detailed investigation of the bifurcation structure of time-delayed feedback induced complex dynamics in a broad-area VCSEL with a saturable absorber. Using bifurcation analysis and direct numerical simulations we have shown that the feedback impacts the homogeneous lasing solution and the localized CS solutions in a similar way, causing multistability of the stationary solutions as well as oscillatory dynamics with either a period doubling or quasiperiodic route to chaos. We have demonstrated that this multistability is caused by a feedback induced phase bifurcation of the stationary solution. The threshold of the phase bifurcation was obtained by a combination of analytical and numerical path continuation methods. The similarity between the bifurcation scenarios of the lasing homogeneous solutions and the CS solutions allows us to perform a complete mapping of the saddle-node, the Andronov-Hopf, period doubling, secondary Hopf (Torus) and the cyclic fold of periodic orbits bifurcations in a plane of the feedback parameters, namely the phase and strength of the feedback.


\section*{Acknowledgements}

    C.S. acknowledges the financial support of project COMBINA (TEC2015-65212-C3-3-P MINECO/FEDER UE). K.P. and M.T. acknowledge the financial support of Interuniversity Attraction Poles program of the Belgian Science Policy Office, under grant IAP P7-35 (photonics@be). K.P. acknowledges Methusalem foundation for financial support. M.T. is a Senior Research Associate with the Fonds de la Recherche Scientifique F.R.S.- FNRS, Belgium. S.G. acknowledges the support of Center of Nonlinear Science (CeNoS) of the University of M\"{u}nster.

\bibliographystyle{apsrev4-1}

%


\end{document}